\documentstyle[prb,aps,twocolumn]{revtex}

\begin{document}

\draft
\twocolumn[\hsize\textwidth\columnwidth\hsize\csname @twocolumnfalse\endcsname

\title{Projection of plane-wave calculations into atomic orbitals}

\author{Daniel Sanchez-Portal, Emilio Artacho, and Jose M. Soler}
\address{
   Instituto de Ciencia de Materiales Nicol\'as Cabrera and
   Departamento de F\'{\i}sica de la Materia Condensada, C-III
\\
   Universidad Aut\'onoma de Madrid, 28049 Madrid, Spain.}
\date{ To appear in Solid State Communications }
\maketitle

\begin{abstract}
The projection of  the eigenfunctions obtained in standard plane-wave
first-principle electronic-structure calculations into atomic-orbital basis
sets is proposed as a formal and practical link between the methods based on
plane waves and the ones based on atomic orbitals. Given a candidate atomic
basis, ({\it i}) its quality is evaluated by its projection into the plane-wave
eigenfunctions, ({\it ii}) it is optimized by maximizing that projection,
({\it iii}) the associated tight-binding Hamiltonian and energy bands are
obtained, and ({\it iv}) population analysis is performed in a natural way.
The proposed method replaces the traditional trial-and-error procedures of
finding appropriate atomic bases and the fitting of bands to obtain
tight-binding Hamiltonians. Test calculations of some zincblende
semiconductors are presented.

\end{abstract}

\pacs{ {\bf Keywords}: D. electronic band structure; A. semiconductors}

]

   Plane waves and atomic orbitals provide two widely used methods for
electronic structure calculations, each of them showing advantages and
disadvantages. This paper proposes a way to use the advantages of each
method to complement the other.

   On one hand, plane waves (PW), in addition to {\it
ab initio} pseudopotentials\cite{hamann} provide a very successful
scheme\cite{cohen} to calculate the ground state properties of a large
number of systems giving accurate results for electron densities,
total energies, atomic forces, geometries, and energy bands. The origin of its
wide applicability is the flexibility, simplicity, and accuracy of the
plane-wave basis set: it has a homogeneous and universal completeness
controlled by a single parameter, the energy cut-off.
It is a floating basis, independent of
atomic positions, and this greatly facilitates the calculation of atomic
forces. Being equally complete everywhere, however, makes the PW basis
inefficient, requiring a relatively large basis size for a given accuracy.
In addition, plane waves require periodic boundary conditions. When the
system is not periodic (molecules, surfaces, defects) such a
periodicity must be imposed artificially by the use of supercells, what
further increases the inefficiency of the basis.

  On the other hand, methods based on the linear combination of atomic
orbitals (LCAO) are far more efficient in terms of basis size, since
atomic orbitals are much better suited to represent the molecular or
Bloch wavefunctions.\cite{eschrig} In addition, LCAO bases have recently
experienced a renewed interest because of their suitability for order-N
methods\cite{ordern} in which computational effort scales linearly with
system size. Local bases also offer a natural way of quantifying magnitudes
like atomic charge, orbital population, bond charge, charge transfer, etc.,
through population analysis.\cite{mulliken} These quantities are essential for
the chemical analysis of a condensed matter system since most of the chemical
language is based on them. They provide a deep physical insight.

  The price to be paid for the advantages mentioned above is the greater
difficulty in the choice of the LCAO basis: since the results of calculations
are affected by
the election of the basis, a testing of the completeness and quality of the
basis is critical. This task is much worse defined than in the case of plane
waves since there are many more parameters than just a plane-wave
energy cutoff, and usually ends up in a very cumbersome and non systematic
trial-and-error procedure.

   Thus, due to the difficulties and ambiguities of the LCAO basis construction
and their less well controlled completeness, they are frequently considered
more approximate than plane waves, and they have been traditionally used
in less accurate tight-binding type of calculations. However, once a well
optimized LCAO basis has been obtained, there is no reason why highly
accurate {\it ab initio} calculations cannot be performed
with it.\cite{abinitiolcao} This rises the question of whether
plane-wave calculations can be used to find an optimal or near-optimal LCAO
basis, which can then be used in more complex situations while keeping a
feasible computational effort.

   Atomic orbitals constitute non-orthogonal basis sets in molecules and
solids. Even though this kind of bases are more cumbersome to handle than
orthogonal ones, they have shown to be more transferable and less
environment-dependent, and to have a shorter range of interactions,
being thus more adequate for non-parametrized LCAO methods.\cite{porfy,eafy}
Non-orthogonalized atomic basis sets are therefore used throughout this
work.\cite{emilio}

   Most schemes proposed to optimize LCAO local bases are based on two
procedures: ({\it i}) the minimization of total energies in atoms, molecules,
or solids,\cite{poirier} and ({\it ii}) the optimization of energy bands as
compared with experiments or with plane-wave calculations.\cite{eschrig,chadi}
In contrast, the present work links the plane-wave and LCAO methods by
projecting the eigenstates of a plane-wave calculation into the Hilbert
space spanned by the atomic basis. This provides an efficient and practical
analytical tool which is applied to ({\it i}) characterize and optimize
atomic basis sets, ({\it ii}) obtain LCAO bands and Hamiltonians, and
({\it iii}) analyze electronic structures by means of population
analysis.\cite{LMTO}

   In this work the reference plane-wave calculations are performed within
the local density approximation (LDA) for electron exchange and
correlation,\cite{lda} using pseudopotentials that replace core
electrons.\cite{TM} These approximations are, however, not essential
to the method. They can be replaced by others at will, the only key factor
being the expansion of the one-particle wave functions in plane waves.
The theory presented in this paper considers the PW reference calculations
to be converged in plane-wave energy cutoff. The explicit consideration of
the uncompleteness of the PW basis will be the subject of a future study.

   Given the PW results for the system of interest, and given the set of
Hamiltonian eigenstates to be considered for the projection, the quality
of an atomic basis is quantified by its ability to represent those eigenstates,
i.e., by measuring how much of the subspace of the Hamiltonian eigenstates
falls outside the subspace spanned by the atomic basis. For that we define the
{\it spilling\cite{spill}} parameter ${\cal S}$
\begin{equation}
\label{desviation}
{\cal S}={1\over N_k} {1\over N_\alpha} \sum_{k}^{N_k}
\sum_{\alpha=1}^{N_\alpha} \langle \psi_\alpha ({\bf k}) | (1-P({\bf k}))
| \psi_\alpha ({\bf k}) \rangle
\end {equation}
where $|\psi_\alpha ({\bf k}) \rangle $ are the PW calculated eigenstates,
and $N_k$ and $N_\alpha$ are the number of calculated {\bf k} points in the
Brillouin zone and the number of bands considered, respectively. $P({\bf k})$
is the projector operator into the subspace of Bloch functions of wave vector
{\bf k}, generated by the atomic basis, and defined as usual for non-orthogonal
basis,\cite {emilio}
\begin {equation}
\label{projector}
P({\bf k})= \sum_{\mu} |\phi_\mu({\bf k}) \rangle \langle \phi^\mu({\bf k})| =
\sum_{\mu \nu} |\phi_\mu({\bf k}) \rangle S^{-1}_{ \mu \nu} ({\bf k})
\langle \phi_\nu({\bf k}) | ,
\end {equation}
where
\begin{equation}
S_{ \mu \nu}({\bf k})=\langle \phi_\mu({\bf k})|\phi_\nu({\bf k}) \rangle
\end{equation}
is the overlap matrix of the atomic basis, and the vectors
$|\phi^\mu({\bf k}) \rangle$ are the dual of the atomic basis, that
satisfy
\begin{equation}
\langle \phi^\mu({\bf k})|\phi_\nu({\bf k}) \rangle =
\langle \phi_\mu({\bf k})|\phi^\nu({\bf k}) \rangle =
\delta_{\mu \nu}.
\end{equation}
${\cal S}$ measures the difference between the plane-wave eigenstates
$|\psi_\alpha ({\bf k}) \rangle $ and their projection into the
atomic-basis $P({\bf k})|\psi_\alpha ({\bf k}) \rangle $. More
precisely, it gives the average of $|| (1-P)|\psi_\alpha ({\bf k}) \rangle
||^2$ over the eigenstates considered for the projection.
The spilling ${\cal S}$ varies between 0 and 1. ${\cal S}=0$ means that
the LCAO projected wave-functions reproduce the PW eigenfunctions exactly.
${\cal S}=1$ means that the basis is orthogonal to the Hamiltonian eigenstates.

  As mentioned above, the quality of an LCAO basis is evaluated not only for a
particular system, but also for a particular set of Hamiltonian eigenstates.
An LCAO basis will reproduce with very different accuracy the
different eigenstates of a Hamiltonian. The choice of eigenstates relevant
to the definition of ${\cal S}$ depends on the particular application.
For ground-state properties of the system (total energy, geometry, electronic
density, chemical analysis) the occupied eigenstates are the ones
to be considered. For quasiparticle excitations or optical properties,
some of the lowest empty states should be also included in the projection.
As shown later, the latter case typically requires a larger atomic basis.

  In addition to the analysis and possible selection of atomic basis sets,
the methodology presented in this work allows for their systematic
optimization in a straightforward way. Given a type of atomic functions
(say Slater-type orbitals, Gaussian-type orbitals) that depend on certain
parameters (say exponents) the problem of optimizing the basis becomes
minimizing ${\cal S}$ as a function of those parameters. Note that
the minimization requires evaluations of ${\cal S}$ for different values
of the parameters, but {\it for a single first-principles calculation}.
Taking into account that the evaluation of ${\cal S}$ represents a minimal
computational effort compared with the actual PW calculation, it is obvious
that this optimization procedure is much more convenient than the direct
trial-and-error procedures using self-consitent LCAO codes.

   Using this procedure we have performed a systematic study comparing
different kinds of atomic orbitals in different solids. This will be presented
in detail elsewhere.\cite{mentira} In that analysis it is found that for many
properties (spilling, charge density and populations, energy bands) the
type of atomic orbitals that gives better results is the following: take the
eigenfunctions of the free atoms using the same basic approximations as for
the solid (LDA, same pseudopotential \cite{TM}) and scale them down by scale
factors $\lambda_l$ (one scale factor per different atomic orbital),
that is, $ \phi_l({\bf r}) = N_l(\lambda_l)\phi_l^{atom}(\lambda_l{\bf r})$,
being $N_l(\lambda_l)$ the normalization factor.
The optimum scale factors are determined by minimization of ${\cal S}$.
Typically the orbitals compress by a small amount except for the orbitals
which are empty in the atomic calculation that compress substantially.
These optimized orbitals are the ones used in the following, except stated
otherwise.

   A further utility of the projection technique is its suitability for
obtaining and analyzing LCAO Hamiltonians (tight-binding parameters) and
their associated energy bands.\cite{chadi} The matrix elements of the
projected Hamiltonian
\begin {equation}
\label {hamiltonian}
H^{LCAO}_{\mu \nu}({\bf k})= \langle \phi_{\mu}({\bf k})|H^{PW}|
\phi_{\nu}({\bf k}) \rangle
\end {equation}
can be obtained using fast Fourier transform algorithms and other techniques
of the PW method:
\begin {equation}
\label {hamiltonian2}
H^{LCAO}_{\mu \nu}({\bf k})=\!\!\!\!\!\!\!\sum_{|{\bf k+G}|^2<E_{max}}
\!\!\!\!\!\!\!\!\!\! \langle
\phi_{\mu}({\bf k})|{\bf k + G}\rangle\langle{\bf k + G}|H^{PW}|
\phi_{\nu}({\bf k}) \rangle
\end {equation}
where $\bf G$ are reciprocal lattice vectors and $E_{max}$ is an energy
cut-off independent of the one used in the PW
calculation, that must be large enough to guarantee a reliable representation
of the atomic basis. It has to be stressed that the proposed procedure
obtains the Hamiltonian matrix elements from first-principles, no free
parameters being fitted.

    The LCAO Hamiltonian matrix is obtained {\it directly} in Bloch space.
By construction, it includes the information of the matrix elements between
localized functions {\it up to infinite neighbors}. This is a very interesting
feature of our method for the characterization of a basis in terms of
its associated band structure: it conveniently separates the effect
of the uncompleteness of the basis from the effect of neglecting matrix
elements beyond some scope of neighbors. These two sources of error are
usually mixed together in the traditional ways of choosing atomic orbitals
from their LCAO bands. With our procedure each approximation can
be analyzed separately.

    To obtain the real-space Hamiltonian matrix elements (tight-binding
parameters) from the reciprocal-space Hamiltonian, we perform an inverse
Bloch transformation
\begin {equation}
\label {parameter}
H^{LCAO}_{\mu \nu}({\bf R}_{\mu \nu})=\sum_{\bf k} H^{LCAO}_{\mu \nu}
({\bf k}) e^{i{\bf k}({\bf R}_\mu - {\bf R }_\nu)}
\end {equation}
where normalization factors which depend on the overlaps are omitted for
clarity. The sum has to be extended to a sufficient number of
{\bf k} points, taking into account that the number of points depends
on the real space range of the interactions.\cite{monkhorst}
For silicon and the STO-4G basis used below, interactions are important up
to third nearest neighbours and a few tens of {\bf k} points have proven to be
enough.

   In Figure 1 (a) and (b) we present the energy bands of silicon for an $sp$
minimal basis and for an $spd$ basis, respectively, with interactions up to
infinite neighbors, as obtained from the projection. They are compared
with the PW energy bands, that were calculated with a well converged plane-wave
basis set. The atomic basis was generated and optimized as explained above,
the $d$ orbital being obtained from an excited configuration of the silicon
atom. The spilling of charge (considering the sum over occupied states in
Eq. 1) is ${\cal S}=0.0076$ and ${\cal S}=0.0007$ for the $sp$ and $spd$ basis,
respectively. The figure shows that the $sp$ basis reproduces the valence band
much better than the lowest conduction bands, giving a rather poor
description of the band gap. The inclusion of the $d$ orbitals substantially
improves both the valence band and the band gap, especially the latter.

    The atomic orbitals used above are relatively extended. For some purposes
a reduction of their extension is desirable. Figure 2 shows the effects of a
contraction of the basis. The energy of the X point rises, showing a tendency
to establish a direct band gap, in addition to the expected narrowing of the
bands and the overall increase of the gap. The shape and the ordering of the
conduction bands is very sensitive to the contraction of the basis. See for
example the change in ordering of the $s$ and $p$ conduction states at $\Gamma$
with the change in scale factor. The charge spillings are ${\cal S}=0.0076$ and
${\cal S}=0.0292$ for the optimum and the contracted basis, respectively.

    Even though the energy bands become worse with an overlocalization of
the atomic orbitals, this can still be interesting if matrix elements for
neighbors beyond a range are wanted to be negligible. The effect of this kind
of approximation is shown  in Figure 3, where the energy bands for silicon are
presented for an STO-4G $sp$ minimal basis\cite{poirier} with interactions up
to first (a), second (b), and third (c) nearest neighbors. They result from
the diagonalization of the LCAO Hamiltonian matrix obtained in Eq. 5, and
they are compared with the result of infinite neighbors. We have observed that
the atomic basis obtained from the scaled
eigenfunctions of the atom usually require a longer range of interactions.
The use of STO-4G orbitals reduces the range of non-negligible interactions.
However, the quality of the bands (infinite neighbors) for the STO basis
(dotted lines in Fig. 3) is worse than for the other basis, showed with
solid lines in Fig. 1 (a), as compared with the plane-wave results, shown
as dotted lines in Fig. 1. Notice that the optimized basis used for
Figure 1 (a) correctly gives an indirect band gap, which is not usually
obtained with standard minimal $sp$ basis sets like the one used in Fig. 3.
The charge spilling of the STO-4G basis is ${\cal S}=0.0513$, considerably
larger than for the basis used in Fig. 1 (a) (${\cal S}=0.0076$).
In LCAO calculations the extension of the atomic orbitals has to be chosen
compromising between the quality required for the band-structure and
valence properties, and the possibility of neglecting matrix elements
beyond a range. The choice of basis depends on the particular application.

   Finally, we show the utility of the projection technique for the analysis
of the results of PW calculations by means of LCAO population analysis. We
use the one proposed by Mulliken.\cite{mulliken} The analysis is performed
on the occupied eigenstates projected into the subspace of the basis.
It has to be done with care since the projected
eigenstates $| \chi_\alpha ({\bf k})\rangle = P({\bf k})|\psi_\alpha
({\bf k})\rangle$ are not orthonormal (the deviation from orthonormality is
related with the charge spilled in the projection\cite{spill}). Proper
consideration of their associated overlap $R_{\alpha \beta}({\bf k})=\langle
\chi_\alpha ({\bf k})| \chi_\beta ({\bf k})\rangle$ is necessary to ensure
charge conservation in the population analisys. This is accomplished
by defining the density operator as
\begin {equation}
\label{density operator}
\hat \rho = \sum_{{\bf k}} \sum_{\alpha}^{occ} |\chi_\alpha({\bf k})\rangle
\langle \chi^\alpha ({\bf k}) | \, ,
\end {equation}
where $|\chi^\alpha ({\bf k})\rangle= \sum_{\beta} R_{\beta \alpha}^{-1}
({\bf k})| \chi_\beta ({\bf k})\rangle$ represent the vectors of the dual
set of the projected eigenstates. The LCAO density matrix is then written
in terms of the dual LCAO basis:
\begin {equation}
\label{density matrix}
{\cal P}_{\mu \nu}=\langle \phi^{\mu}|\hat \rho|\phi^{\nu} \rangle \, ,
\end {equation}
the charge associated to an orbital $\mu$ then being
\begin {equation}
\label{charge of an orbital}
Q_\mu=\sum_\nu {\cal P}_{\mu \nu } S_{\nu \mu} .
\end {equation}

   Results of this procedure are shown in Table I, where charge
transfers are calculated for different heteropolar zincblende materials.
They are compared with results of selfconsistent LCAO calculations
showing a remarkable agreement. The basis functions used for these
population analysis are the scaled atomic orbitals discussed above,
which showed to be best suited for the purpose.

   In summary, it has been shown how the simple projection of the eigenstates
obtained from PW calculations into the space spanned by an LCAO basis can be
useful for multiple purposes: (i) to evaluate and optimize LCAO basis sets,
(ii) to analyze LCAO band structures, (iii) to obtain tight-binding parameters,
and (iv) to perform population
analysis. The procedures are straightforward and more systematic than
the previous methods to perform equivalent tasks. The main idea of this paper
could also be used for the improvement of first-principles algorithms.
This possibility is presently being explored.


{\it Acknowledgments} - We acknowledge useful discussions with
F. Yndur\'ain. E.~A. also acknowledges discussions with S.~G.~Louie.
This work has been supported by the Direcci\'on General de Investigaci\'on
Cient\'{\i}fica  y Tecnol\'ogica of Spain (DGICYT) under grant PB92-0169.

%
%

%


\begin{figure}
\caption{
Electronic band structure of silicon calculated projecting the plane-wave LDA
Hamiltonian into an $sp$ atomic basis (a), and into an $spd$ atomic basis (b),
both shown in solid lines. Dotted lines show the plane-wave LDA band structure.
The basis functions are obtained as discussed in the text. }
\label{sp-spd}
\end{figure}
\begin {figure}
\caption {
Effect of different contractions of an $sp$ atomic basis in the silicon band
structure. Solid lines are for optimum scale factors and dotted lines for a
further contraction of the basis functions of a 1.2 factor.
The basis functions are obtained as discussed in the text. }
\label {2contractions}
\end{figure}
\begin{figure}
\caption{
Band structure of silicon as a function of the range of non-neglected
interactions:
interactions up to first (a), second (b), and third (c) nearest neighbors.
In dotted lines the result for infinite neighbors is shown as reference
(in (c) they are hardly distinguishible from the solid lines).
The atomic basis functions are standard STO-4G taken from Ref. 10. }
\label{neighb}
\end{figure}
%

%
%
\begin{table}
\caption{
Calculated charge transfer for some zincblende semiconductors. $Q_C$ and
$Q_A$ stand for the valence charge on the cation and the anion, respectively,
$\delta Q$ for the charge transfer with respect to neutral atoms, and
${\cal S}$ for the spilling. Numbers in parenthesis were obtained from
Hartree-Fock LCAO calculations,$^{18}$
except for GaAs, for which LDA LCAO was used.$^{19}$ }

\vspace{8pt}
\begin{tabular}{llllll}

& Basis & ${\cal S}$  & $Q_C$ & $Q_A$ & $\delta Q$ \\
\hline
BN   & $s,p$   & 0.0022 & 2.19 (2.14) & 5.81 (5.86) &  0.81  (0.86) \\
BP   & $s,p$   & 0.0038 & 3.51 (3.34) & 4.49 (4.66) & --.51 (--.34) \\
AlP  & $s,p$   & 0.0035 & 2.15 (2.20) & 5.85 (5.80) &  0.85  (0.80) \\
SiC  & $s,p$   & 0.0071 & 2.30 (2.19) & 5.70 (5.81) &  1.70  (1.81) \\
GaAs & $s,p$   & 0.0041 & 2.58        & 5.42        &  0.42         \\
     & $s,p,d$ & 0.0010 & 2.78 (2.88) & 5.22 (5.12) &  0.22  (0.12) \\
\end{tabular}
\label{charges}
\end{table}


\begin{references}
\bibitem{abinitiolcao} LCAO is the most frequently used type of basis in the
very accurate first-principles calculations in quantum chemistry. See Ref.
\onlinecite{poirier} for references.
\bibitem{spill} If projecting the occupied eigenstates, the defined
spilling parameter gives the proportion of the total charge lost (spilled)
when projecting.
\end{references}
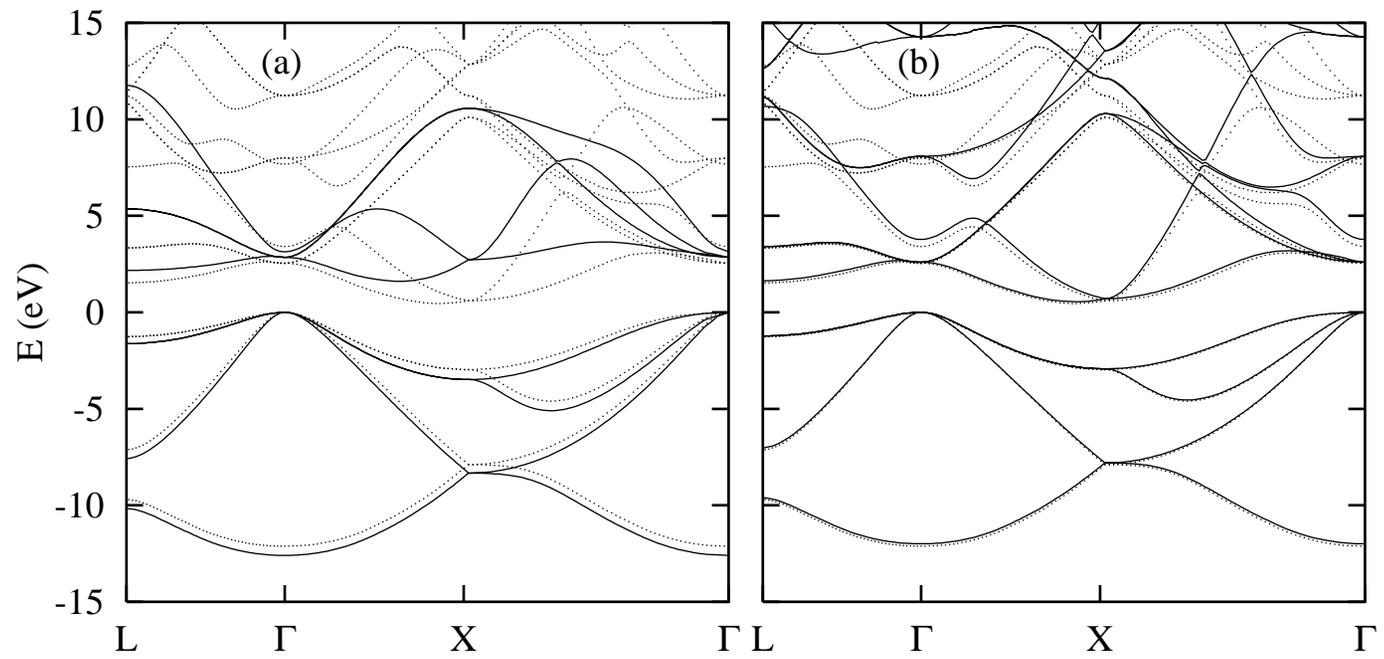
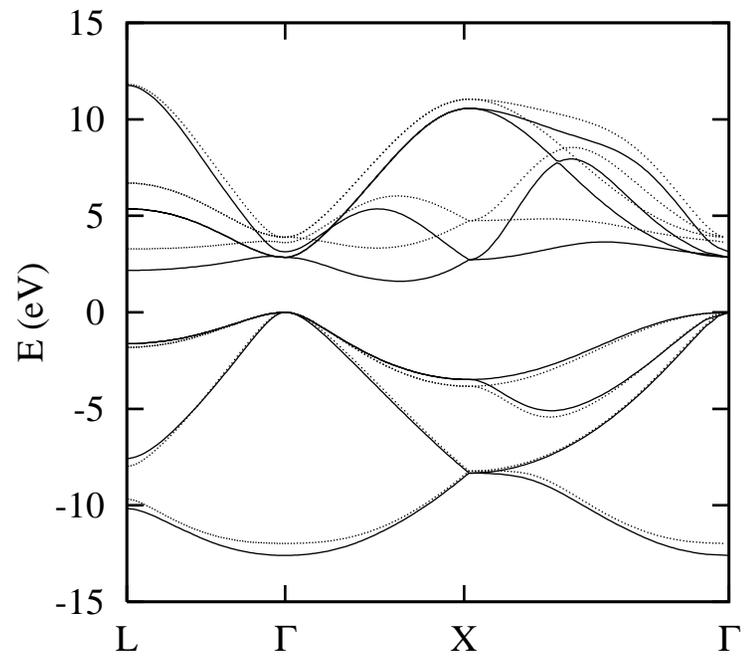
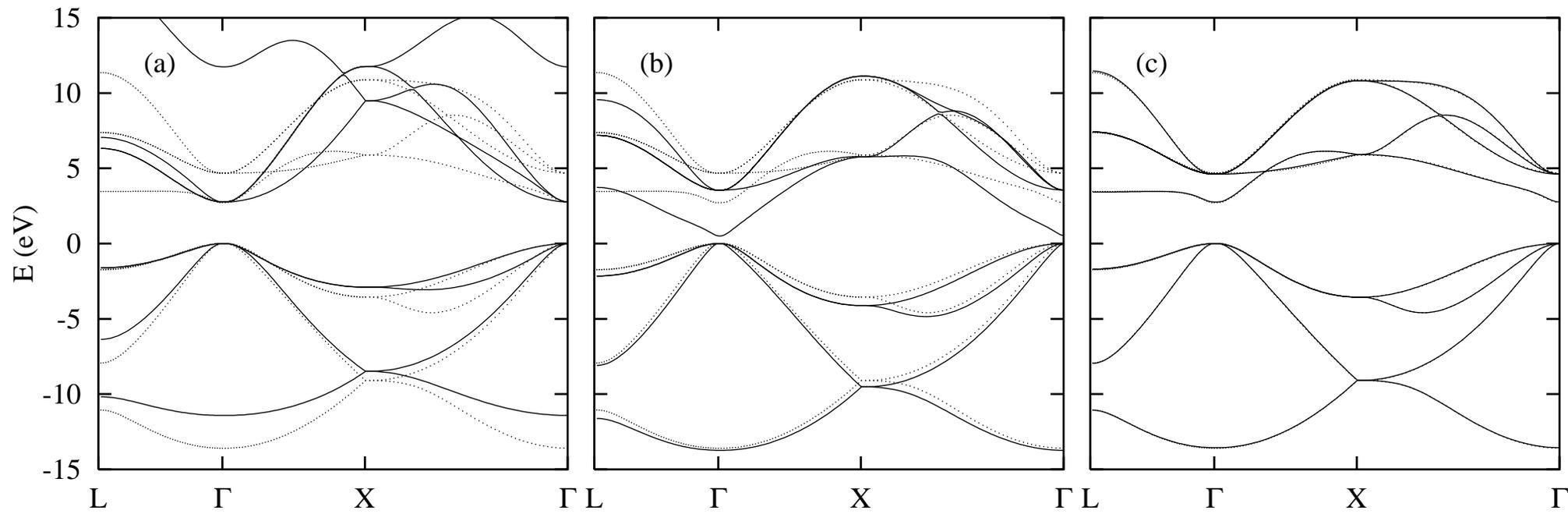
\end{document}